# GRChombo: An adaptable numerical relativity code for fundamental physics

**Tomas Andrade[1], Llibert Areste Salo[2], Josu C. Aurrekoetxea[3], Jamie Bamber[4], Katy Clough[4], Robin Croft[5], Eloy de Jong[3], Amelia Drew[5], Alejandro Duran[6], Pedro G. Ferreira[4], Pau Figueras[2], Hal Finkel[7], Tiago França[2], Bo-Xuan Ge[3], Chenxia Gu[2], Thomas Helfer[8], Juha Jäykkä[9], Cristian Joana[9], Markus Kunesch[5], Kacper Kornet[5], Eugene A. Lim[3], Francesco Muia[5], Zainab Nazari[10, 11], Miren Radia[5], Justin Ripley[5], Paul Shellard[5], Ulrich Sperhake[5, 12], Dina Traykova[4], Saran Tunyasuvunakool[5], Zipeng Wang[8], James Y. Widdicombe[3], and Kaze Wong[8]**

**1** Departament de Fisica Quantica i Astrofisica, Institut de Ciencies del Cosmos, Universitat de Barcelona, Marti i Franques 1, 08028 Barcelona, Spain **2** School of Mathematical Sciences, Queen Mary University of London, Mile End Road, London E1 4NS, United Kingdom **3** Theoretical Particle Physics and Cosmology, King's College London, Strand, London, WC2R 2LS, United Kingdom **4** Astrophysics, Oxford University, Denys Wilkinson Building, Keble Road, Oxford OX1 3RH, United Kingdom **5** Department of Applied Mathematics and Theoretical Physics (DAMTP), University of Cambridge, Centre for Mathematical Sciences, Wilberforce Road, Cambridge CB3 0WA, United Kingdom **6** Intel Iberia, Torre Picasso Plaza Pablo Ruiz Picasso 1 Madrid, 28020 Spain **7** Argonne National Laboratory (ANL), 9700 S. Cass Avenue, Argonne, IL 60439-4815, United States **8** Henry A. Rowland Department of Physics & Astronomy, Johns Hopkins University, 3701 San Martin Drive, Baltimore, Maryland (MD) 21218, United States **9** Cosmology, Universe and Relativity at Louvain (CURL), Institut de Recherche en Mathematique et Physique, University of Louvain, 2 Chemin du Cyclotron, 1348 Louvain-la-Neuve, Belgium **10** Department of Physics, Bogazici University, 34342 Bebek, 80820 Istanbul, Turkey **11** HECAP Section, Abdus Salam International Centre for Theoretical Physics (ICTP), 34151, Trieste, Italy **12** California Institute of Technology, Pasadena, California 91125, USA

## Summary

The 2015 detection of gravitational waves (GWs) from a binary black hole merger (B. P. Abbott & others, 2016) was a breakthrough moment for science. More detections have since been made by the Advanced LIGO/Virgo network (Aasi & others, 2015; B. P. Abbott & others, 2018; R. Abbott & others, 2020; Acernese & others, 2015) and future ground and space based detectors (Amaro-Seoane & others, 2017; Hu & Wu, 2017; Luo & others, 2016; Saleem & others, 2021; Somiya, 2012) will further expand our reach.

Strong gravity regimes are described by the *Einstein Field Equation* (EFE) of General Relativity (Einstein, 1916)

$$R_{\mu\nu} - \frac{1}{2}Rg_{\mu\nu} = 8\pi G T_{\mu\nu} \,, \tag{1}$$

where $g_{\mu\nu}$ is the gravitational metric describing spacetime distances, $R$ and $R_{\mu\nu}$ are related to its second derivatives in space and time, and $T_{\mu\nu}$ is the stress-energy tensor of any matter or fields present. Analytic solutions to the EFE only exist where there is a high degree of symmetry; in general the equations must be solved numerically. The need for observational predictions has thus led to the development of *numerical relativity* (NR), methods for numerically solving the above equations, typically utilising high-performance computing (HPC) resources. Expanding out the tensorial notation above, the EFE is a set of coupled, nonlinear second-order partial differential equations for $g_{\mu\nu}$, which describes the curvature of spacetime in the presence of stress-energy $T_{\mu\nu}$, that is, schematically the equation we are trying to solve has the form:

$$\partial_t \partial_t g_{\mu\nu} \sim \partial_x \partial_x g_{\mu\nu} + \partial_y \partial_y g_{\mu\nu} + \partial_z \partial_z g_{\mu\nu} + \text{nonlinear cross terms} + 8\pi G T_{\mu\nu} \ , \quad (2)$$

where the indices $\mu, \nu$ run over the spacetime indices – in 4 dimensions, $t, x, y, z$. Given that $g_{\mu\nu}$ is symmetric in its indices, this gives a set of ten coupled nonlinear partial differential equations, sourced by the stress-energy of any matter or fields present in the spacetime.

One common approach to NR is to specify an initial spatial distribution for the metric and matter fields (subject to certain constraints), and then solve a time evolution for all metric and matter quantities, thus populating their values thoughout the four-dimensional spacetime. The canonical example of this is the simulation of two black holes in orbit around each other, which permits extraction of the gravitational wave signal produced during the merger. Such numerical results have been instrumental in discovering signals in the noisy LIGO/VIRGO detector data, as well as confirming the predictions of GR to a high precision in the strong field regime (B. P. Abbott & others, 2016; R. Abbott & others, 2020).

GRChombo is an open-source code for performing such NR time evolutions, built on top of the publicly available Chombo software (Adams & others, 2015) for the solution of PDEs. Whilst GRChombo uses standard techniques in NR, it focusses on applications in theoretical physics where adaptivity, both in terms of grid structure, and in terms of code modification, are key drivers.

## Key features of GRChombo

Since its initial announcement in 2015 (Clough et al., 2015), the GRChombo code has become a fully mature, open-source NR resource.

The key features of GRChombo are as follows:

- BSSN/CCZ4 formalism with moving punctures: GRChombo evolves the Einstein equation in the BSSN (Baumgarte & Shapiro, 1999; Nakamura et al., 1987; Shibata & Nakamura, 1995) or CCZ4 (Alic et al., 2012; Gundlach et al., 2005) formalism with conformal factor $\chi = \det(\gamma_{ij})^{-1/3}$, where $\gamma_{ij}$ is the induced metric on the spatial hyperslices. Singularities of black holes are managed using the moving puncture gauge conditions (Baker et al., 2006; Campanelli et al., 2006), and Kreiss-Oliger dissipation (Kreiss et al., 1973) is used to control high-frequency noise, both from truncation and the interpolation associated with regridding.

- Boundary Conditions: The code implements periodic, Sommerfeld (radiative), extrapolating and reflective boundary conditions.

- Initial Conditions: The current examples provide analytic or semi-analytic initial data for black hole binaries, Kerr black holes and scalar matter. The code also incorporates a standalone version of the TwoPunctures code (Ansorg et al., 2004) for accurate binary BH data of arbitrary spins (up to the usual limit for Bowen-York data of around $a/M = 0.9$ for the dimensionless spin parameter), masses and momenta.

- Diagnostics: GRChombo has routines for finding black hole horizons, calculating spacetime masses, angular momenta, densities, fluxes and extracting gravitational waves.

- C++ class structure: GRChombo is written in the C++ language, and makes heavy use of object oriented programming (OOP) and templating.

- Parallelism: GRChombo uses hybrid OpenMP/MPI parallelism with explicit vectorisation of the evolution equations via intrinsics, and is AVX-512 compliant. Our code demonstrates efficient strong scaling up to several thousand CPU-cores for a typical BH binary problem, and further for larger problem sizes.

- Adaptive Mesh Refinement: The underlying Chombo code provides Berger-Oliger style (M. J. Berger & Oliger, 1984) AMR with block-structured Berger-Rigoutsos grid generation (M. Berger & Rigoutsos, 1991). The tagging of refinement regions is fully flexible and can be based on truncation error or other user-defined measures.

The code continues to be actively developed with a number of ongoing projects to add new features.

## Statement of Need

Several 3+1D NR codes using the moving puncture formulation already exist and are under active development. The Einstein Toolkit (http://einsteintoolkit.org/), with its related Cactus (http://cactuscode.org) (Loffler & others, 2012; Schnetter et al., 2004), and Kranc (http://kranccode.org) (Husa et al., 2006) infrastructure used by LEAN (Sperhake, 2007; Zilhao et al., 2010) and Canuda (https://bitbucket.org/canuda) (Witek et al., 2019). Other notable but non-public codes include BAM (Bruegmann et al., 2008; Marronetti et al., 2007), AMSS-NCKU (Galaviz et al., 2010), PAMR/AMRD and HAD (East et al., 2012; Neilsen et al., 2007). Codes such as SPeC (Pfeiffer et al., 2003) and bamps (Hilditch et al., 2016) implement the generalised harmonic formulation of the Einstein equations using a pseudospectral method, and discontinuous Galerkin methods are used in SpECTRE (https://spectre-code.org) (Deppe et al., 2021; Kidder & others, 2017) (see also (Cao et al., 2018)). NRPy (http://astro.phys.wvu.edu/bhathome) (Ruchlin et al., 2018) is a code aimed for use on non-HPC systems, which generate C code from Python, and uses adapted coordinate systems to minimise computational costs. CosmoGRaPH (https://cwru-pat.github.io/cosmograph) (Mertens et al., 2016) and GRAMSES (Barrera-Hinojosa & Li, 2020) are among several NR codes targeted at cosmological applications (see (Adamek et al., 2020) for a comparison) and which also employ particle methods. Simflowny (https://bitbucket.org/iac3/simflowny/wiki/Home) (Palenzuela et al., 2018), like CosmoGRaPH, is based on the SAMRAI infrastructure, and has targeted fluid and MHD applications. GRAthena++ (Daszuta et al., 2021) makes use of oct-tree AMR to maximise scaling.

While GRChombo is not the only open-source NR code, its unique features (detailed above) have made it one of the premier codes for numerical relativity, especially in the study of fundamental physics beyond standard binary mergers. In particular, GRChombo's highly flexible adaptive mesh refinement scheme allows for complicated "many-boxes-in-many-boxes" topology , enabling users to simulate non-trivial systems, such as ring-like configurations Helfer, Aurrekoetxea, et al. (2019) and inhomogeneous cosmological spacetimes (Aurrekoetxea, Clough, et al., 2020; Clough et al., 2017; Clough, Flauger, et al., 2018; Joana & Clesse, 2021). Nevertheless, with its efficient scalability and AMR capabilities, it can also play a leading role in the continuing efforts to simulate "standard" binary mergers to the required sensitivities required for the upcoming LISA space mission (Radia et al., 2021). Finally, GRChombo's object-oriented and template-based code can be rapidly modified for non-standard problems such as higher-dimensional spacetimes (Andrade et al., 2020; Bantilan et al., 2019; Figueras et al., 2016, 2017), modified gravity systems (Figueras & França, 2020) and additional fundamental fields (Alexandre & Clough, 2018; Bamber et al., 2021; Clough, Dietrich, et al., 2018; Clough et al., 2019; Dietrich et al., 2019; Helfer et al., 2017; Helfer, Lim, et al., 2019; Muia et al., 2019; Nazari et al., 2021; Widdicombe et al., 2020).

## Key research projects using GRChombo

The wide range of fundamental physics problems for which the code has been used so far includes:

- the simulation of inhomogeneous pre-inflationary spacetimes, bubble collisions and pre-

heating in early universe cosmology (Aurrekoetxea, Clough, et al., 2020; Clough et al., 2017; Clough, Flauger, et al., 2018; Joana & Clesse, 2021).

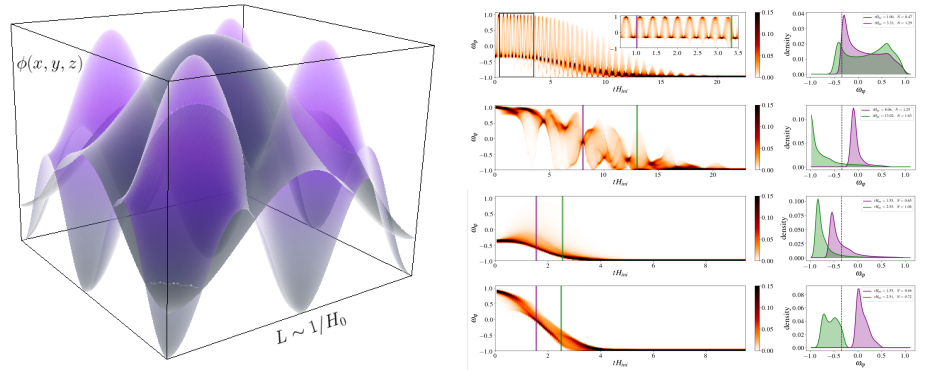

*Images of Imhomogeneous inflaton field in (Aurrekoetxea, Clough, et al., 2020) and evolution of the equation of state and density in (Joana & Clesse, 2021).*

- the study of modified gravity, and violation of cosmic censorship (Andrade et al., 2020; Bantilan et al., 2019; Figueras et al., 2016, 2017; Figueras & França, 2020).

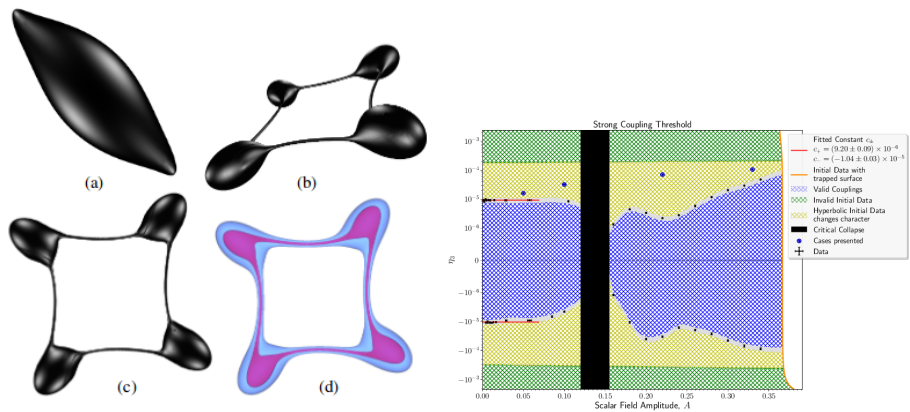

*Images of testing cosmic censorship with higher-dimensional black rings in (Figueras et al., 2017) and mapping regions of validity for modified gravity in (Figueras & França, 2020).*

- the formation, collapse and collisions of exotic compact objects (ECOs) and dark matter stars (Clough, Dietrich, et al., 2018; Dietrich et al., 2019; Helfer et al., 2017; Helfer, Lim, et al., 2019; Muia et al., 2019; Nazari et al., 2021; Widdicombe et al., 2018, 2020).

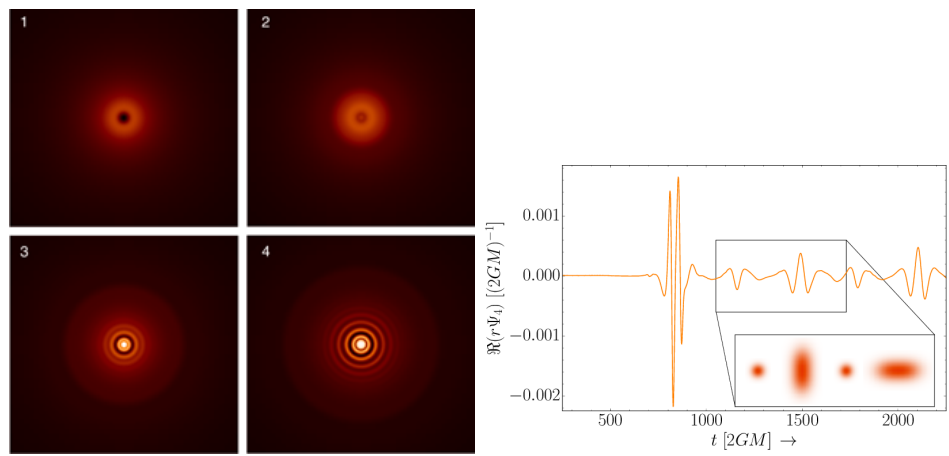

*Images of axion star collapse from (Helfer et al., 2017) and GW signals from an ECO collision (Helfer, Lim, et al., 2019).*

- gravitational wave emission from cosmic string collapse (Aurrekoetxea, Helfer, et al., 2020; Helfer, Aurrekoetxea, et al., 2019) and scalar radiation from global cosmic(/axion) strings (Drew & Shellard, 2019).

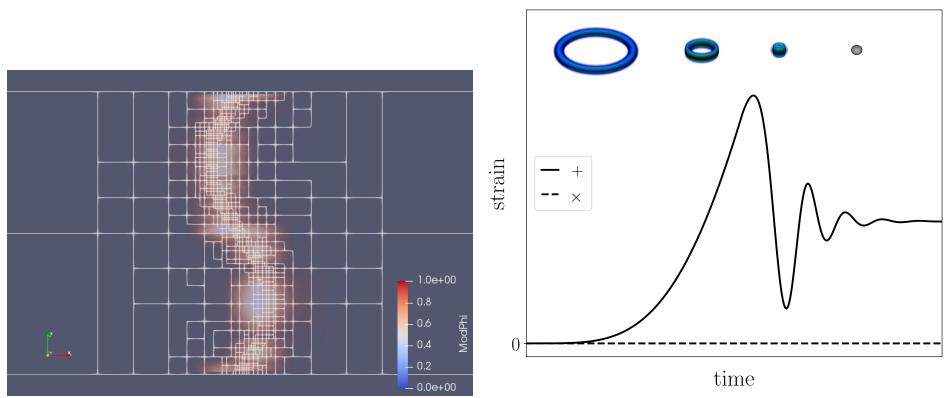

*Images of global axion strings from (Drew & Shellard, 2019) and the GW signal from cosmic string loop collapse in (Aurrekoetxea, Helfer, et al., 2020).*

- the study of light bosonic dark matter and neutrino-like particles in black holes environments (Alexandre & Clough, 2018; Bamber et al., 2021; Clough et al., 2019; Traykova et al., 2021).

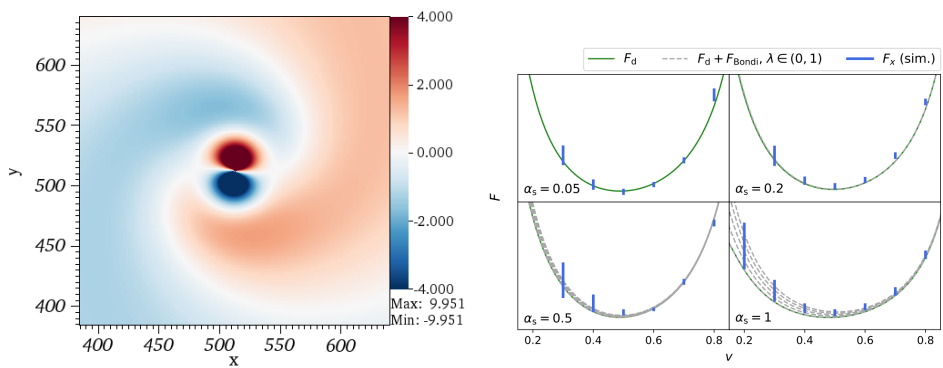

*Images of scalar field accretion around a spinning BH from (Bamber et al., 2021), and the relativistic scaling of dynamical friction from (Traykova et al., 2021).*

- the study of gravitational recoil in unequal mass binaries (Radia et al., 2021).

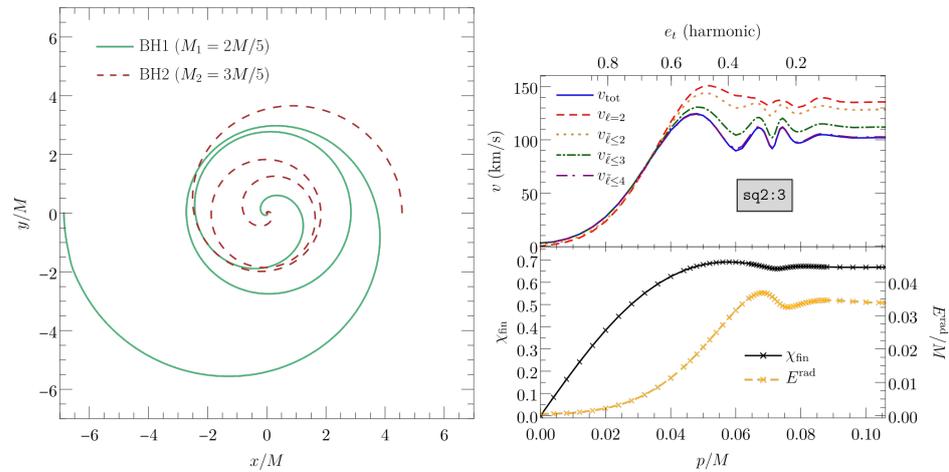

*Images of the BH trajectories and recoil velocities, spin and radiated energy in (Radia et al., 2021).*

## Acknowledgements

The GRChombo collaboration gratefully acknowledges support to its members by the ERC, UKRI/STFC, Intel, The Royal Society, PRACE and DiRAC.

In particular, P Ferreira, JB, KC and DT acknowledge support from the European Research Council (ERC) under the European Union's Horizon 2020 research and innovation programme (grant agreement No 693024). P Figueras also acknowledges support from the ERC for the project ERC-2014-StG 639022-NewNGR and a Royal Society University Research Fellowship, Grant No. UF140319, URF/R/201026 and RGF/EA/180260. TF acknowledges support under the Royal Society grant RS/PhD/181177. AD has been supported by an EPSRC iCASE Studentship in partnership with Intel (EP/N509620/1, Voucher 16000206) and is currently supported by a Junior Research Fellowship at Homerton College, Cambridge. F.M. is funded by a UKRI/EPSRC Stephen Hawking fellowship, grant reference EP/T017279/1. This work has been partially supported by STFC consolidated grant ST/P000681/1. Z.N. is supported by ICTP-Sandwich Training Educational Programme (STEP). HF was supported by the U.S. Department of Energy (DOE) Office of Science under Contract DE-AC02-06CH11357.

We thank the developers of the Chombo code for their assistance and guidance on using their code, and the Intel Parallel Computing Centre at the University of Cambridge for their support of our code development. We acknowledge the support of the Intel Visualization team, led by Jim Jeffers, notably the collaboration on in-situ visualization with Carson Brownlee.

GRChombo users have benefited from the provision of HPC resources from:

- DiRAC (Distributed Research utilising Advanced Computing) resources under the projects ACSP218, ACSP191, ACTP183 and ACTP186. Systems used include:

  - Cambridge Service for Data Driven Discovery (CSD3), part of which is operated by the University of Cambridge Research Computing on behalf of the STFC DiRAC HPC Facility (www.dirac.ac.uk). The DiRAC component of CSD3 was funded by BEIS capital funding via STFC capital grants ST/P002307/1 and ST/R002452/1 and STFC operations grant ST/R00689X/1. DiRAC is part of the National e-Infrastructure.

  - DiRAC Data Intensive service at Leicester, operated by the University of Leicester IT Services, which forms part of the STFC DiRAC HPC Facility (www.dirac.ac.uk). The equipment was funded by BEIS capital funding via STFC capital grants ST/K000373/1 and ST/R002363/1 and STFC DiRAC Operations grant ST/R001014/1. DiRAC is part of the National e-Infrastructure.

- DiRAC at Durham facility managed by the Institute for Computational Cosmology on behalf of the STFC DiRAC HPC Facility (www.dirac.ac.uk). The equipment was funded by BEIS capital funding via STFC capital grants ST/P002293/1 and ST/R002371/1, Durham University and STFC operations grant ST/R000832/1. DiRAC is part of the National e-Infrastructure.

- DIRAC Shared Memory Processing system at the University of Cambridge, operated by the COSMOS Project at the Department of Applied Mathematics and Theoretical Physics on behalf of the STFC DiRAC HPC Facility (www.dirac.ac.uk). This equipment was funded by BIS National E-infrastructure capital grant ST/J005673/1, STFC capital grant ST/H008586/1, and STFC DiRAC Operations grant ST/K00333X/1. DiRAC is part of the National e-Infrastructure.

- DiRAC Complexity system, operated by the University of Leicester IT Services, which forms part of the STFC DiRAC HPC Facility (www.dirac.ac.uk ). This equipment is funded by BIS National E-Infrastructure capital grant ST/K000373/1 and STFC DiRAC Operations grant ST/K0003259/1. DiRAC is part of the National e-Infrastructure.

- PRACE (Partnership for Advanced Computing in Europe) resources under grant numbers 2018194669, 2020225359. Systems used include:

  - SuperMUCNG, Leibniz Supercomputing Center (LRZ), Germany

  - JUWELS, Juelich Supercomputing Centre (JSC), Germany

  - Cartesius (SURF), Netherlands

  - Marenostrum (BSC), Spain

- the Argonne Leadership Computing Facility, including the Joint Laboratory for System Evaluation (JLSE), which is a U.S. Department of Energy (DOE) Office of Science User Facility supported under Contract DE-AC02-06CH11357.

- the Texas Advanced Computing Center (TACC) at the University of Austin HPC and visualization resources URL: http://www.tacc.utexas.edu, and the San Diego Supercomputing Center (SDSC) URL: https://www.sdsc.edu, under project PHY-20043 and XSEDE Grant No. NSF-PHY-090003

- Consortium des Équipements de Calcul Intensif (CÉCI), funded by the Fonds de la Recherche Scientifique de Belgique (F.R.S.-FNRS), Belgium

- the Marconi HPC resources and software support (awarded by CINECA), Italy

- the Glamdring cluster, Astrophysics, Oxford, UK

- The Fawcett cluster, Faculty of Mathematics, Cambridge, UK

- the Argo cluster at ICTP, Trieste, Italy

- the Apocrita cluster at QMUL, UK

- The Athena cluster at HPC Midlands Plus, UK

- The Cosmo cluster at CURL, University of Louvain, Belgium